\documentclass[pss]{wiley2sp} % provides new 2008 pss two-column style (no alternative manuscript style output available at present)
\usepackage{amsmath}
\usepackage{amssymb}
\usepackage{bm}              % uncomment these two packages if you
\newcommand{\bsigma}{{\bm \sigma}}

\newcommand{\ve}{\varepsilon}
\newcommand{\hh}{\hat{H}}

\begin{document}

% Title of the article
\title{Role of non-collinear magnetization: from ferromagnetic nano wires to quantum rings}

% Abbreviated title for the page headers
\titlerunning{Non-collinear magnetization}

% Authors
\author{%
  Nicholas Sedlmayr\textsuperscript{\Ast,\textsf{\bfseries 1}},
  Vitalii Dugaev\textsuperscript{\textsf{\bfseries 2,3}},
  Jamal Berakdar\textsuperscript{\textsf{\bfseries 1}}}

% Abbreviated list of authors for the page headers
\authorrunning{Nicholas Sedlmayr et al.}

%E-mail-address of corresponding author
\mail{e-mail
  \textsf{sedlmayr@physik.uni-kl.de}%, Phone:
 % +xx-xx-xxxxxxx, Fax: +xx-xx-xxx
 }

% author's affiliations/addresses
\institute{%
  \textsuperscript{1}\,Department of Physics, Martin-Luther-Universit\"at Halle-Wittenberg,
Heinrich-Damerow-Str. 4, 06120, Halle, Germany\\
  \textsuperscript{2}\,Department of Physics, Rzesz\'ow University of Technology,
al. Powsta\'nc\'ow Warszawy 6, 35-959 Rzesz\'ow, Poland\\
  \textsuperscript{3}\,Department of Physics and CFIF, Instituto Superior T\'ecnico,
TU Lisbon, av. Rovisco Pais, 1049-001, Lisbon, Portugal}

\received{XXXX, revised XXXX, accepted XXXX} % do not change, will be filled in by the publisher
\published{XXXX} % do not change, will be filled in by the publisher

%Please select four to six PACS-codes from the enclosed list (PACS.txt) or from www.aip.org/pacs)
 % For example: 71.20.Ps

\abstract{We consider interaction effects related to a nonuniform magnetization in ferromagnetic nanowires and their possible generalization to nanorings.
%We consider interaction effects related to a nonuniform magnetization in ferromagnetic nanowires and nanorings.
First we show that the electric current in a ferromagnetic nanowire with
more than one domain wall
induces an exchange coupling between the walls mediated by the spin-dependent
interference of scattered carriers.
This interaction reveals a complex behavior
as a function of mutual orientations and separation of the domain walls, thus affecting
the domain wall dynamics.
%Then we consider the behavior of domain walls in a magnetized quantum ring.
Then we consider how the theory should be modified in the case of a magnetized quantum ring.
In this case the situation is more complicated because
the vortex and onion magnetization states in the ring are not truly ferromagnetic.}

\maketitle   % please do not remove

\section{Introduction}

Domain walls (DWs) are noncollinear magnetization regions separating
areas of \emph{different} homogenous magnetizations. In nanostructures they are the subject
of numerous investigations and are also important for applications \cite{yamanouchi04,yamaguchi04,parkin08}.
In low dimensional nanostructures, such as the wires and rings we are here concerned with, the coupling between
the electrons and the DWs is enhanced in comparison with the bulk scenario \cite{klaui08}.
This can drastically alter the transport properties of the nanostructures \cite{ebels00,chopra02,ruster03}.
The spin flip of the carriers as they traverse the region of non-collinearity leads to a
spin-torque acting on the DW, and consequently to the possibility of current-induced
DW motion \cite{yamanouchi04,yamaguchi04}.

DWs can exist not only in magnetic nanowires but can also be created in magnetic
nanorings \cite{vaz07,klaeui08} between the
regions of a locally defined homogenous magnetization. Fully ferromagnetic rings, where the magnetization direction
is globally preserved, are not experimentally found, though they inform the majority of theoretical considerations.
Instead, the magnetization direction usually follows the shape of the annulus\footnote{This can be explained by simple
magnetostatics accounting for the demagnetization fields.}.
In this case it is clear that the DWs always come in pairs and as such a natural question to ask is how they
interact, mediated by the electrons in the system. As a precursor to this question we discuss such
interactions, and their effect on the DWs in nanowires \cite{sedlmayr09,sedlmayr10}. This will form the bulk of the results
presented in this paper. The case of quantum dot with a non-collinear magnetic order has been
discussed in \cite{sedlmayr08}. Before addressing the case of magnetic rings, we discuss several
fundamental points. In addition to the
effects of electrons on the DWs we are also interested in how the ferromagnetic nature of the system and
the presence of the DWs affect such well-known phenomena as the Aharonov-Bohm effect \cite{aharonov59,webb85},
persistent currents \cite{buettiker83,levy90}, and conductance \cite{gefen84}.

It is established that strong carrier scattering and interference
results in  long-range interactions
between impurities on metal surfaces.
The question of whether and how the carriers' spin dependent scattering
mediates interactions between DWs is addressed
here. We have found that the DWs are coupled thusly: due to the scattering from the first DW
a spiral spin density wave is created. This then acts as a spatially non-uniform torque on
the second DW whose energetically stable shape and position
therefore has a non-uniform dependence on the distance from the first DW. This
is different from the ordinary spin-torque transfer
in bulk spin valve systems \cite{slonczewski96,berger96,barnas05}
or magnetic tunnel junctions \cite{slonczewski89,theodonis06} insofar as in this case the  DWs
spatial arrangement, in addition to the magnetization direction, is current controlled.

There are a number of experimental works on permalloy ferromagnetic rings \cite{vaz07,klaeui08}.
In these works the existence of the Aharonov-Bohm effect was experimentally
confirmed \cite{kasai02,sekiguchi08}
and the studies of the magnetoresistance effect have also been presented \cite{klaeui03,lai03,jain08}.
However, it is still not clear how the variation of magnetization in these systems
affects the state and dynamics of the DWs. Though our aim is to consider DWs in the nanoscale rings,
we discuss first the rings without DWs. The main question is how the vortex state
in a magnetic ring affects its electric properties.  In particular, how the corresponding
edge states look
like and how persistent currents and Aharonov-Bohm type effects are modified by choosing
different magnetization configuration.

The first question to answer involves analyzing the
electron wavefunctions of the ferromagnetic ring \cite{tan96,kim99}.
%Following this idea we hope to generalize the system to the onion state, though most probably only
%considering a one-dimensional ring at this juncture.  such a system has domain walls and so
%our previous interests will return.
Note that in the following we call the magnetization of ring "ferromagnetic"
%because of course a true
%ferromagnetic configuration of the ring would have a uniform magnetization pointing in the
%same direction everywhere (c.f.~Tatara, et al.\cite{tatara01}).  In our case, and indeed
%the experimental systems,
when the magnetization vector is directed along the $\theta $ angular unit vector in the cylindric
coordinates of the ring. In this configuration there is no energy cost from the demagnetization field,
unlike the true ferromagnetic case.  The vortex state instead incurs the smaller penalty of
not having exactly parallel magnetic moments everywhere, though locally they will be approximately parallel for
a large ring. Note also that as we are dealing with a ring, there is no energy cost due to the vortex core.

The Zeeman splitting of the electron spectrum in a magnetic field has been included in
Refs.~\cite{tatara01,tatara04}.
The overall effect of this, as far as the Aharonov-Bohm effect is concerned, is to alter the
dephasing length scales.  The spin-orbit coupling has also been included but only
perturbatively where it mixes the spin channels. Though the spin-orbit coupling could be considered
analogous to what we wish to calculate (which also couples angular orbital momentum and spin)
we solve this model exactly.

Another interesting idea concerns the magnetically textured mesoscopic ring of Ref.~\cite{loss90}.
In this
model, one considers the inhomogeneous magnetic field distorted from the $z$-direction around
the ring, like a crown. This induces a form of spin-orbit coupling which can lead to
persistent currents even in the absence of a magnetic flux threading the ring. The method
can be useful for decoupling the orbital and spin
degrees of freedom by using a Feynman path integral representation.
As far as the domain wall motion is concerned some efforts have been also made for
semi-rings \cite{meier07}.
%but not for the full ring structure and its possible effects.
%Nonetheless the first task remains a proper description of the ferromagnetic ring itself.
%
It has been also proposed to take into account the inhomogeneous anisotropy, which arises
naturally in magnetic nanoring \cite{bruno05,dugaev07,dugaev07}. It was shown that the
effect of the resulting Berry phase of electrons is quite similar to that of an external field affecting
the equilibrium magnetization profile.

\section{Wires}

\subsection{Theoretical model}

\begin{figure}
\includegraphics[height=0.45\textwidth,angle=90]{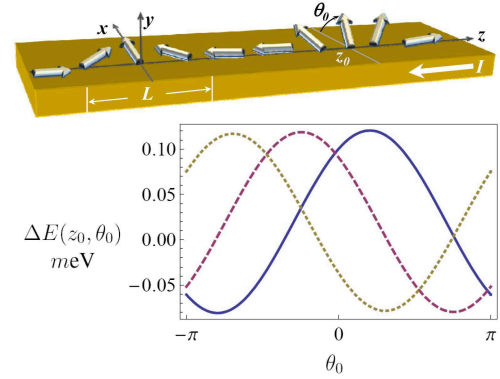}
\caption[$\Delta E$]{(Color online) Top panel:
A schematic showing the DWs magnetization profile (thick arrows).
 $L$ is the DW width, $z_0$ and $\theta_0$ are respectively
  the DW position and orientation with respect to the DW at $z=0$, $I$ is the current direction.
 Lower panel: Interaction energy $\Delta E(z_0,\theta_0)$ as
a function of $z_0$ and $\theta_0$.
The solid  curve is for $z_0=30$~nm, the dashed curve is for $z_0=37.5$~nm,
and the dotted curve is for $z_0=45$~nm. }
\label{deltae2d}
\end{figure}

We consider first a long magnetic wire with two DWs, down which a current, $I$, is passed
(as shown schematically in Fig.~\ref{deltae2d}). Assuming that the distance  $z_0$ between
DWs is larger than the phase coherence length $L_\phi$, we can consider them as two independent
scatterers, but for $z_0\lesssim L_\phi$  the current transmission mediates
DW coupling. For definiteness,
we assume that one of the DWs (located at $z=0$) is pinned, e.g. by a geometric constriction,
and concentrate on the effect
of the current on the second DW, initially (i.e., for $I=0$) located  at $z=z_0$. For
$I=0$ each DW has  an extension $L$. We model the one-dimensional wire with the following
Hamiltonian, $\bar H$, of noninteracting electrons  coupled (with a coupling constant $J$)
to a spatially non-uniform
magnetization (DW) profile ${\bf M}(z)$:
\begin{eqnarray}
\bar H=\int dz\, a_\alpha ^\dag (z)
\left[-\frac{\delta _{\alpha\beta}\partial _z^2}{2m}\,
-J\, \bm{\sigma}_{\alpha\beta}\cdot {\bf M}(z)\right] a_\beta (z),
\end{eqnarray}
(we use units with $\hbar=1$). Here $a^\dag_{\alpha}$ and $a_\alpha $ are the creation and
annihilation operators of electrons with spin $\alpha $. We use a local unitary transformation $T(z)$ \cite{kor,tatara,dugaev1} (often referred to as a gauge transformation) which transforms the second
term with
nonuniform magnetization into the constant Zeeman splitting term and an additional
spin-dependent spatially varying
potential $U_{\alpha\beta}(z)$. For $k_F L\gtrsim 1$
this potential can be treated perturbatively \cite{kor,tatara,dugaev1,ijmp} ($k_F$
is the Fermi wave vector) whereas the sharp domain walls, i.e. when $k_F L< 1$, require a
different formalism \cite{dugaev3}.

The matrix $T(z)$ is defined by
$T^\dag(z)\, {\bm \sigma} \cdot {\bf n}(z)\, T(z)=\sigma ^z$,
where ${\bf n}$ is the unit vector along ${\bf M}$,
${\bf M}(z)=M\, {\bf n}(z)$.
Then the transformed Hamiltonian  $H=T^\dag(z)\, {\bar H} \, T(z)
$ is
\begin{eqnarray}
\label{hamiltonian}
H=\int dz\, a_\alpha^\dag(z)\left[-
\frac{\delta_{\alpha\beta}\partial _z ^2}{2m}\,
+U_{\alpha\beta}(z)
-JM\sigma^z_{\alpha\beta}\right ] a_\beta(z).\nonumber
\end{eqnarray}\begin{eqnarray}{}\end{eqnarray}
For a wire with two DWs we can parameterize
the magnetization profile by two angles $\varphi (z)$ and $\theta (z)$
(cf. Fig.~\ref{deltae2d})
\begin{eqnarray}
{\bf n}(z)&=&\big( \cos \theta \, \sin \varphi ,\;
\sin \theta \, \sin \varphi ,\;
\cos \varphi \big) ,\label{magnetization1}\\
\varphi(z)
&=&\underbrace{\cos^{-1}\big(\tanh\big[\frac{z}{L}\big]\big)}_{=-\varphi_1(z)} +\underbrace{\cos^{-1}\big(\tanh\big[\frac{z-z_0}{L}\big]\big)}_{=-\varphi_2(z)}
\label{magnetization2}
\end{eqnarray}
(for details see Ref.~\cite{sedlmayr09}). The relative orientation between
the two DWs (situated, respectively, at $z=0$ and $z=z_0$) is set by
angle $\theta(z)$. We set
$\theta_1 =0$ at the first DW  and $\theta_2 =\theta_0$ around the second one
(see Fig.~\ref{deltae2d}). In the following we consider the case of $z_0> L$, which can
be treated perturbatively as the coupling of the DWs is relatively small. Then we can write
$U(z)\approx U_1(z)+U_2(z)$, where ($j=1,2$)
\begin{eqnarray}
U_j(z)&=&
\frac{[\varphi_j'(z)]^2}{8m}+i\sigma ^y
\bigg[\frac{\varphi_j''(z)}{4m}+\frac{\varphi_j'(z)\, \partial_z}{2m}\bigg]
\cos \theta_j(z)
\nonumber\\&&
-i\sigma ^x \bigg[\frac{\varphi_j''(z)}{4m}+\frac{\varphi_j'(z)\, \partial_z}{2m}\bigg]
\sin \theta_j(z).
\label{u2}
\end{eqnarray}
Note that this approach is generalizable to any number of DWs which are sufficiently far apart.

As shown in Refs.~\cite{tatara,dugaev1,ijmp}, for a single DW and $k_FL\geq 1$ a
perturbative approach is appropriate for treating the electron scattering  from the
DW potential (\ref{u2}).
Thus, assuming $\psi ^0(z)$ to be the wave function of electron with energy $\varepsilon $
in the wire without potential (\ref{u2}), we
find the first-order correction
due to the perturbation $U_1(z)$, i.e. due to scattering from the first DW, as
\begin{eqnarray}
\delta \psi _\ve (z)=\int_{-\infty}^\infty dz'\,
G_\ve(z,z')\, U_1(z')\, \psi ^0(z').
\end{eqnarray}
The Green's function $G_\varepsilon (z,z')$ corresponds to the unperturbed
Hamiltonian (2) with $U(z)=0$. It is diagonal in the spin space with elements
\begin{eqnarray}\label{freeg}
G_{\ve \sigma }(z,z')=-\frac{im}{k_\sigma }\, e^{ik_\sigma |z-z'|},
\end{eqnarray}
where $k_\sigma \approx k^0_\sigma +\frac{i}{2\tau_\sigma}\frac{m}{k^0_\sigma}$
for lifetimes $\tau_\sigma \gg \ve_F^{-1}$, and
$k^0_{\uparrow,\downarrow }=\left[ 2m\, (\ve+\mu\pm JM)\right] ^{1/2}$.

The interaction energy of the two DWs due to the single scattered state
$\psi _{\ve \sigma }(z)=\psi ^0_{\ve \sigma }(z)+\delta \psi _{\ve \sigma }(z)$ is
\begin{eqnarray}
\Delta E_\sigma =\int_{-\infty}^\infty dz\,
\delta\psi^\dagger_{\ve\sigma}(z)\,
U_2(z)\, \delta\psi_{\ve\sigma}(z).
\end{eqnarray}
To find the total interaction energy we should sum up the contributions of all scattering
states in the energy range
between $\ve_F$ and $\ve_F+e\Delta\phi /2$, for an applied voltage
$e\Delta\phi/2\ll \ve_F$. Then we find the current-induced coupling of the DWs:
\begin{eqnarray}
\Delta E=\frac{e\Delta\phi}{\sqrt{2}\pi}
\bigg(\frac{\Delta E_\uparrow }{v_\uparrow }
+\frac{\Delta E_\downarrow}{v_\downarrow }\bigg),
\label{deltaeeq}
\end{eqnarray}
where $v_\sigma =k^0_\sigma /m$ is the velocity of spin up and down electrons
at the Fermi level.

For numerical calculations we use the parameters of magnetic semiconductors
\cite{ruster03,sugawara08}. It should be noted that in the case of metallic nanowires
the 1D limit is also achievable \cite{claessen}.
Here we take the parameters as in Ref.~\cite{ruster03}, i.e.
$\lambda_F=6$~nm; an effective mass of $m=0.5m_e$
($m_e$ is free electron mass); $L=\lambda_F$; $JM=15$~meV; $\ve_F=83.7$~meV;
and $e\Delta\phi=0.1\ve_F$.
We also assume relatively large mean free path of $l=500$~nm, corresponding to essential
ordering of Mn ions.
The width of the wall may be as small as the atomic size in the
presence of constrictions \cite{bruno99,pietzsch00,ebels00}, hence well below the DW
lengths in bulk materials.
In such a situation, the interaction between the DWs increases due to strongly enhanced
DW scattering \cite{jb_du,jb_du2,jb_du3,jb_du4}.
The interaction energy shown in Fig.~\ref{deltae2d} depends periodically on the DWs
relative angle $\theta_0$ and distance $z_0$. It
results in an oscillating motion of the DW along the axis $z$
as well as an oscillating direction of DW polarization. In its turn this will have an
effect on the spin torque driving the DW dynamics, which we will now focus on.

We find the spin density due to a single transmitted electron wave of spin
$\sigma$ is
\begin{eqnarray}
{\bf S}_\sigma (z)
=\psi ^\dag _{\ve\sigma}(z)\, T(z)\,
\bm{\sigma}\, T^\dag (z) \, \psi _{\ve\sigma}(z),
\end{eqnarray}
and the total current-induced spin density is \cite{dugaev2}
\begin{eqnarray}\label{spin}
{\bf S}(z)=\frac{e\phi}{2\pi}
\bigg( \frac{{ \bf S}_{\uparrow}}{v_\uparrow }
+\frac{{\bf S}_{\downarrow }}{v_\downarrow } \bigg) .
\end{eqnarray}
We find that the correction to the spin density follows
the magnetization profile with additional Friedel oscillations,
which are a superposition of two waves with periods $k^{-1}_{F\uparrow }$ and
$k^{-1}_{F\downarrow }$. The oscillations in the spin density are smaller in
magnitude than the overall spin density profile and decay with increasing $z$.

\begin{figure}
\includegraphics[width=0.45\textwidth]{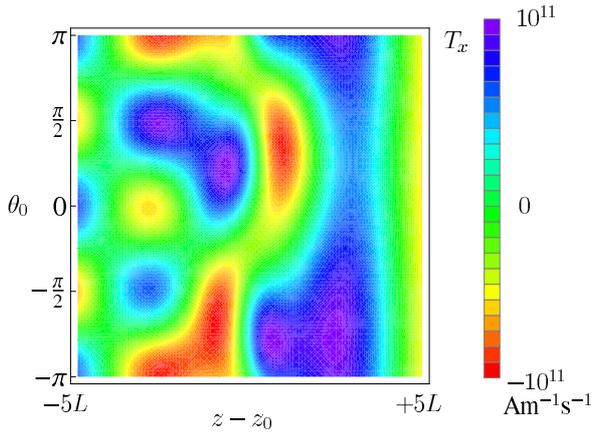}
\caption[$T_x(z,\theta_0)$: Spin Torque Density Plot Correction]
{(Color online) The $x$-component
of the current-induced spin torque, as defined in Eq.~\ref{torque},
acting at the second domain wall as a function of $z$ and $\theta_0$.
}\label{densitytorx}
\end{figure}

We calculate the current-induced torque acting on the magnetization at $z$, for the
second DW located at $z_0$, from
\begin{eqnarray}
\label{torque}
\Delta {\bf T}(z,z_0,\theta_0)
=-\frac{\gamma J}{\sigma_{cs}}{\bf M}(z,z_0,\theta_0)\times\Delta {\bf S}(z,z_0,\theta_0),
\end{eqnarray}
where
$\gamma=g\mu_B$, $g$ is the Land\'e factor and $\mu_B$ is the Bohr magneton.
We assume a thin nanowire with a cross section  of $\sigma_{cs}=100\times20$~nm${}^2$
as in Ref.~\cite{ruster03}. In Eq.~(15)
$\Delta {\bf S}$ is the correction to the electron spin density due to
scattering.
The calculated torque on the second DW is shown in Fig.~\ref{densitytorx}
for the $x$-component,
where $z_0=50\, L$ and $M\approx5.56\times10^{4}$Am${}^{-1}$ were used \cite{sugawara08}.
The correction to the spin torque shows that the
force upon the DW depends strongly on their relative orientations.

%\begin{figure}
%\includegraphics[width=0.45\textwidth]{contourtorz.eps}
%\caption[$T_x(z,\theta_0)$: Spin Torque Density Plot Correction]
%{(Color online) The $z$-component
%of the current-induced spin torque
%around the second domain wall as a function of $z$ and $\theta_0$.
%}\label{densitytorz}
%\end{figure}

To inspect the  current-induced dynamics of the  DW at $z=z_0$,
we evaluate the accumulated spin density that acts on
the DW at $z=z_0$. The DW magnetization dynamics are then
modeled using the Landau-Lifshitz equation with the effect of magnetic anisotropy and damping:
\begin{eqnarray}
\label{magn}
\partial_t{\bf M}
=-\frac{\gamma J}{\sigma_{cs}}{\bf M}\times {\bf S}[{\bf M}]
+\frac{\gamma K'}{M^2}{\bf M}\times\hat{{\bf x}}M_x
\nonumber\\
+\frac{\alpha}{M}{\bf M}\times \partial_t{\bf M}.
\end{eqnarray}
%
%\begin{figure}
%\includegraphics[width=0.45\textwidth]{magthetapi4.eps}
%\caption[$\vec{M}(t,\theta_0=\pi/4)$:Magnetization]
%{(Color online) The time dependence
%of the magnetization with the initial condition for the second wall to be
%at an angle of $\theta_0=\pi/4$ to the first wall. This is the solution
%to Eq.~\ref{magn}.  The solid
%curve is the $x$-component, dashed  the $y$-component,
%and dotted  the $z$-component. Taken at the centre of the domain wall.}\label{magthetapi4}
%\end{figure}
%
As an initial condition we assume that the magnetization profile in the wire
without electric current is described by Eq.~(\ref{magnetization2}).
We should note that the relative orientation of the
walls at the start of motion does play a role in the type of motion we see.
Here we present it for an arbitrary configuration.

\begin{figure}
\includegraphics[width=0.45\textwidth]{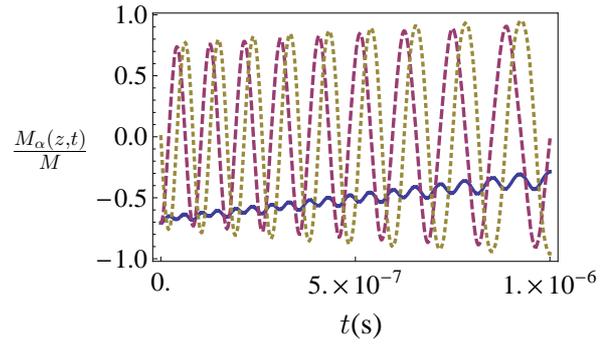}
\caption[$\vec{M}(t,\theta_0=\pi/4)$: Magnetization]{(Color online) The magnetization dynamics of the second DW
when the anisotropy is included, see Eq.~\ref{magn}. The solid
curve is the $x$-component, dashed the $y$-component, and dotted
the $z$-component. Taken at the centre of the domain wall.}\label{anis}
\end{figure}

\begin{figure}
\includegraphics[width=0.45\textwidth]{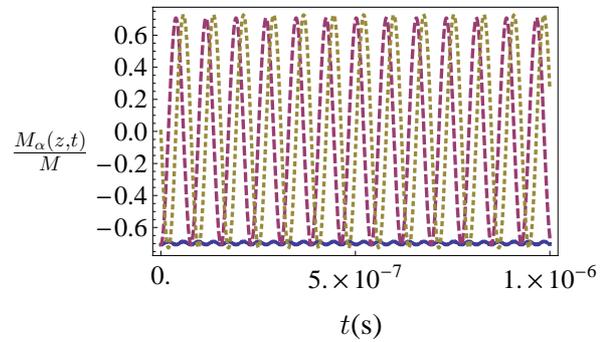}
\caption[$\vec{M}(t,\theta_0=\pi/4)$: Magnetization]{(Color online) The magnetization
dynamics of the second DW without the influence of the first DW,
when the anisotropy is included, see Eq.~\ref{magn}. The solid
curve is the $x$-component, dashed  the $y$-component, and dotted
the $z$-component. Taken at the centre of the domain wall.}\label{aniszero}
\end{figure}

We take the anisotropy constant to be $K'=-10$, and therefore the $x$-axis
to be a hard magnetization axis.  Figure \ref{anis} shows the
effects of anisotropy on the domain wall motion. The anisotropy dampens motion in the
x-direction, thus exacerbating the y and z oscillations.  This is also in contrast to
the case where we ignore the first domain wall. In this case, although the anisotropy
does introduce motion around the centre of the domain wall it does not involve a decaying
$x$-component, see Fig.~\ref{aniszero}.

\begin{figure}
\includegraphics[width=0.45\textwidth]{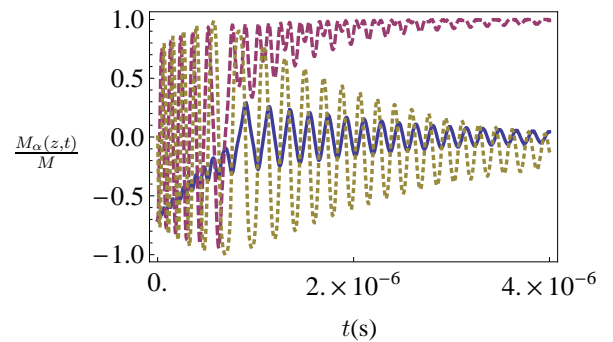}
\caption[$\vec{M}(t,\theta_0=\pi/4)$: Magnetization]{(Color online) The magnetization dynamics of the second DW
when both anisotropy and damping are included, see Eq.~\ref{magn}.  The solid
curve is the $x$-component, dashed  the $y$-component, and dotted
the $z$-component. Taken at the centre of the domain wall.}\label{anidamping}
\end{figure}

Finally let us include weak Gilbert damping also ($\alpha=0.01$), we then find a switch
over between behaviour dominated by the anisotropy and behaviour dominated by the damping.
See figure \ref{anidamping}.

\section{Quantum rings}

\subsection{Model}

The following Hamiltonian
\begin{eqnarray}
\hh=-\frac{1}{2m}\bigg[\frac{1}{r}\partial_r\big(r\partial_r\big)
+\frac{1}{r^2}\bigg(\partial_\theta-i\frac{\phi}{\phi_0}\bigg)^2\bigg]
\nonumber\\-JM\hat{\mathbf{e}}_\theta \cdot \vec{\bsigma}+V(r),
\end{eqnarray}
describes a two dimensional ferromagnetic ring of electrons confined by some
radial potential $V(r)$.
Here we use the usual notations: $J$ is the exchange coupling constant and $m$ is
the electron mass.
In correspondance to the ring geometry it is convenient to use the polar coordinates.
We denote
\begin{eqnarray}
\phi&=&\oint_\gamma\vec{A} \cdot d\vec{r}
=\int\int_S\vec{B}\cdot d\vec{S},
\end{eqnarray}
$\phi_0=h/e$ is the magnetic flux quantum, $\vec{B}$ is the magnetic field and
$\vec{A}$ is the vector potential.
One can simplify Hamiltonian (14) making a local rotation in
spin space to remove the $\theta$ dependence from the $JM$ term.
%, then we can Fourier transform the $\theta$ component.
We use the following unitary transformation:
\begin{eqnarray}\label{rotate}
\mathbf{U}(\theta)=\frac{1}{\sqrt{2}}\begin{pmatrix}i&1\\-e^{i\theta}&-ie^{i\theta}\end{pmatrix}.
\end{eqnarray}
Note that this transformation preserves the same boundary conditions for the wavefunctions.
The transformed Hamiltonian is $\hh'=\mathbf{U}^\dagger(\theta)\hh\mathbf{U}(\theta)$ and
the corresponding wave function is
$\tilde{{\bm\psi}}(r,\theta)=\mathbf{U}^\dagger(\theta){\bm\psi}(r,\theta)$.
We obtain
\begin{eqnarray}
\hh'=-\frac{1}{2m}\bigg[\frac{1}{r}\partial_r\big(r\partial_r\big)
+\frac{1}{r^2}\bigg(\partial_\theta-i\frac{\phi'}{\phi_0}\bigg)^2
\nonumber\\
-\frac{1}{4r^2}-\frac{i}{r^2}{\bm \sigma}_y\bigg(\partial_\theta-i\frac{\phi'}{\phi_0}\bigg)\bigg]
-JM\bsigma_z+V(r),
\end{eqnarray}
where we introduced $\phi'=\phi-\phi_0/2$. The latter means the appearance of an additional
phase of an electron $\phi _0/2$ related to the ring geometry.

The angular dependence can be dealt with by a Fourier transform.
%provided we are in the situation where there are no domain walls.
We use the following definition:
\begin{eqnarray}
{\bm\psi}(r,\theta)&=&\sum_{l=-\infty}^\infty e^{il\theta}
{\bm\psi}_l(r)\, \textrm{ and}\nonumber\\
{\bm\psi}_l(r)&=&\int_0^{2\pi}\frac{d\theta}{2\pi}e^{-il\theta}{\bm\psi}(r,\theta).
\end{eqnarray}
Finally we obtain the following transformed Hamiltonian:
\begin{eqnarray}
\hh'=-\frac{1}{2m}\bigg[\frac{1}{r}\partial_r\big(r\partial_r\big)
-\frac{1}{r^2}{l_\phi'}^2 -\frac{1}{4r^2}+\frac{l_\phi'}{r^2}{\bm \sigma}_y\bigg]
\nonumber\\-JM\bsigma_z+V(r),
\end{eqnarray}
where $l_\phi'=l-\phi'/\phi_0$.
>From this point one can find the appropriate one-dimensional Hamiltonian \cite{meijer02}
which can be further used to calculate the spin current in the ring without the DWs.
%From the full two-dimensional model there are several questions worth considering.
%Principally we can consider how the ferromagnetic magnetization term affects the edge states in the ring.

\subsection{Onions and vortices}

Once we understand the problem of the vortex state without the DWs, we can consider peeling back
the layers of the onion state.  The onion state refers to a ring in which there are two domain
walls present, and hence two regions of oppositely directed magnetization.  One can consider such
state in the limits of sharp and adiabatic DWs (as compared to the electrons' wavelength).
The following refers to the 1D model.
%, and the validity of it should also be
%checked with reference to the model in the previous section, as described above.
Note that one should be
careful in taking the 1D limit when a spin-orbit interaction is present
%and in our case we have some analogous terms present
\cite{meijer02}.

\subsection{Magnetic ring without the domain walls}

In this case we can make the transformation of Eq.~\eqref{rotate}.  Then using
$\hh'=\mathbf{U}^\dagger(\theta)\hh\mathbf{U}(\theta)$ and $\tilde{{\bm\psi}}(\theta)
=\mathbf{U}^\dagger(\theta){\bm\psi}(\theta)$
we obtain
\begin{eqnarray}
\hh'=-\frac{1}{2m\rho^2}\bigg[\partial_\theta-i\frac{\phi'}{\phi_0}\bigg]^2
+\frac{1}{4}\frac{1}{2m\rho^2}\nonumber\\+ i\bsigma_y\frac{1}{2m\rho^2}\bigg[\partial_\theta
-i\frac{\phi'}{\phi_0}\bigg]-JM\bsigma_z.
\end{eqnarray}
After the Fourier transform we diagonalize the Hamiltonian; $\hh''=\mathbf{N}^\dagger(l)\hh'\mathbf{N}(l)$
and $\tilde{\tilde{{\bm\psi}}}(l)=\mathbf{N}^\dagger(l)\tilde{{\bm\psi}}(l)$.
\begin{eqnarray}
\mathbf{N}(l)=\frac{1}{\sqrt{1-{A^+_{l'}}^2}}\begin{pmatrix} 1&iA^+_{l'}\\A^+_{l'}&i\end{pmatrix}.
\end{eqnarray}
$A^+_{l'}=-il_\phi'/(\sqrt{{l_\phi'}^2+\mu^2}-\mu)$ with $\mu=2JMm\rho^2$. Then we obtain the exact formulas for the energy spectrum
\begin{eqnarray}
&&\varepsilon _{l(1,2)}
=\frac{1}{2m\rho^2}\bigg[ {l_\phi'}^2+\frac{1}{4}
\nonumber\\
&&\mp \bigg(\mu
\frac{1+{A^+_{l'}}^2}{1-{A^+_{l'}}^2}-2il_\phi'\frac{A^+_{l'}}{1-{A^+_{l'}}^2}\bigg)\bigg].
\end{eqnarray}
%Note that in this section $m$ is the conjugate of the angle and as such an integer.
%In the case where we have two domain walls present m is a parameter we introduce by hand and
%is just a function of the energy, $\varepsilon$, which takes as of yet unknown values.

\begin{figure}
\begin{center}
\includegraphics[width=0.45\textwidth]{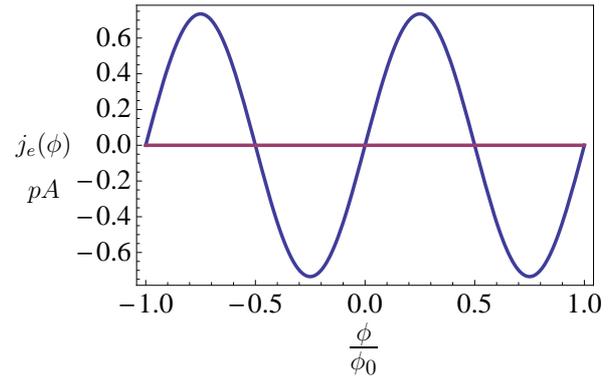}
\caption[Persistent Current]{The persistent current, $j_e(\phi)$, in a one-dimensional
simple ferromagnetic ring.}\label{pecurrent}
\end{center}
\end{figure}

%\begin{figure}
%\begin{center}
%\includegraphics[width=0.45\textwidth]{current.eps}
%\caption[Persistent Current]{The persistent current in a one-dimensional simple ferromagnetic ring.
%The blue curve is the current, $j_e(\phi)$, and the red curve is the ``spin current'', $j_s(\phi)$.}
%\label{pcurrent}
%\end{center}
%\end{figure}

Now we can directly calculate the persistent current. Using
\begin{eqnarray}\label{pc}
j_i(\phi)=-\sum_l\frac{\partial\varepsilon_{li}}{\partial\phi}f(\varepsilon_{li}-\mu)
\end{eqnarray}
(where $i=1,2$) we define the charge current as $j_e(\phi)=j_1(\phi)+j_2(\phi)$.
It is presented in Fig.~\ref{pecurrent} as a function of $\phi $.

\begin{figure}
\begin{center}
\includegraphics[width=0.45\textwidth]{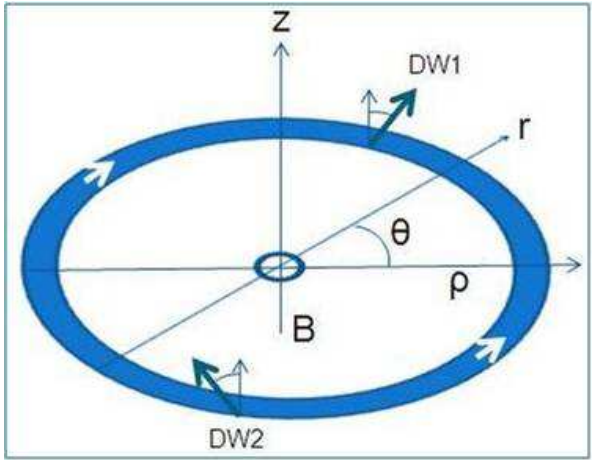}
\caption[Schematic]{A schematic view of a ferromagnetic ring with an applied magnetic field.}
\label{schematic}
\end{center}
\end{figure}

\subsection{The onion state with adiabatic domain walls}

In this section we use a modified version of our model for a wire with two domain walls
\cite{sedlmayr09} shown in Sec.~2.  By mapping the circle onto a line,
and directing the magnetization along $\hat{\mathbf{e}}_\theta$ instead of $\hat{\mathbf{e}}_z$,
we can easily construct a suitable model.  The mapping is given by
\begin{eqnarray}
x(\theta)=\frac{\rho}{L}\tan\bigg(\frac{\theta-\pi}{2}\bigg).
\end{eqnarray}
This mapping distorts the relative sizes of the domain walls depending on where on the ring they
reside,
%nonetheless it should be suitable for us.
and its accuracy increases with increasing the ring diameter.
\begin{eqnarray}
\varphi(\theta)=\underbrace{\pi-\cos^{-1}\big[\tanh[x(\theta)-x(\theta_1)]\big]}_{=\varphi_1(\theta)}+
\nonumber\\
\underbrace{\pi-\cos^{-1}\big[\tanh[x(\theta)-x(\theta_2)]\big]}_{=\varphi_2(\theta)}.
\end{eqnarray}
where $\theta_i$ is the position in the ring of the $i$th wall, which we take to be at an angle $\eta_i$
to the $z$-axis, see Fig.~\ref{schematic}. The magnetization is then
\begin{eqnarray}
\frac{\mathbf{M}(\theta)}{M}=\cos[\varphi(\theta)]\hat{\mathbf{e}}_\theta
+\sin[\eta(\theta)]\sin[\varphi(\theta)]\hat{\mathbf{e}}_r
\nonumber\\
+\cos[\eta(\theta)]\sin[\varphi(\theta)]\hat{\mathbf{e}}_z,
\end{eqnarray}
and the Hamiltonian is given by
\begin{eqnarray}
\hh=-\frac{1}{2m\rho^2}\bigg[\partial_\theta-i\frac{\phi}{\phi_0}\bigg]^2
-J\vec{M}(\theta).\vec{\bsigma}.
\end{eqnarray}
where $\eta(\theta)$ is a function, which takes the value $\eta_i$ around the $i$th wall and slowly interpolates between the walls.

We can locally rotate the spin direction as before. The resultant Hamiltonian,
\begin{eqnarray}
\hh'=-\frac{1}{2m\rho^2}\bigg[\bigg(\partial_\theta-i\frac{\phi}{\phi_0}\bigg)^2 -\frac{1}{4}-i\bigg(\partial_\theta-i\frac{\phi}{\phi_0}\bigg)\bsigma_y\bigg]
\nonumber\\
-J\tilde{\mathbf{M}}(\theta)\cdot \vec{\bsigma},\qquad
\end{eqnarray}
includes the following magnetization profile:
\begin{eqnarray}
\frac{\tilde{\mathbf{M}}(\theta)}{M}
=\cos[\varphi(\theta)]\hat{\mathbf{e}}_z-\sin[\eta(\theta)]\sin[\varphi(\theta)]\hat{\mathbf{e}}_x \nonumber\\+\cos[\eta(\theta)]\sin[\varphi(\theta)]\hat{\mathbf{e}}_y.
\end{eqnarray}
After that we can make a gauge transformation to reduce the magnetization profile to a constant
Zeeman term and an additional potential. However, the non-diagonal portions of the transformed Hamiltonian
complicate significantly the exact calculation of the energy spectrum.

Therefore, we consider a simplified approach which is justified for large magnetic ring, $k\rho \gg 1$,
where $k$ is the electron wave vector along the ring. Using Eq.~(29) we write
\begin{eqnarray}
\hh'=-\frac{1}{2m}\bigg[\bigg(\partial_z-\frac{i\phi}{\rho \phi_0}\bigg)^2
-\frac{1}{4\rho ^2}-\frac{i}{\rho }\bigg(\partial_z-\frac{i\phi}{\rho \phi_0}\bigg)\bsigma_y\bigg]
\nonumber\\
-J\tilde{\mathbf{M}}(\theta)\cdot \vec{\bsigma},\qquad
\end{eqnarray}
where $z=\rho \theta $ is the coordinate along the wire. Due to the periodicity in angle $\theta $,
the condition of periodicity $z\to z+2\pi \rho $ should be preserved.
If $k\rho \gg 1$ the second term in Eq.~(30) is small and can be taken into account perturbatively.
If we neglect the small correction we come to the 1D Hamiltonian, which can be treated using the
same methods as the linear model of Sec.~2. The main difference is that the gauge potential
related to the DWs is an additional correction to the gauge related to the ring curvature and
to the magnetic flux through the ring.

It should be noted that the ring topology allows only an even number of
DWs. In the case of two DWs on the ring, the current-induced interaction
between them can be described by the spin torque formula Eq.~\eqref{torque}. On the
other hand, the persistent current in the magnetic ring, Eq.~\eqref{pc}, would
necessarily induce the current-induced force and an interaction between
the walls, which puts them both into a kind of circular motion. However,
one can also expect that the damping associated with the DW motion would
lead to a slow decay of the persistent current. We can assume that a
slight pumping of power from an external source can make the motion of
DWs on the ring non-decaying, which makes this problem attractive for
possible applications.

%As such the presence of the real ferromagnetic magnetization in this
%system means we can no longer apply our theory for nanowires to this case. It is first necessary
%to find an appropriate way of simplifying the above magnetization so that perturbation theory can be applied.

\section{Summary}

We have shown how the presence of more than one domain wall in a wire affects the domain wall
dynamics and the energy of the electron system. We calculated the energy due to the
current mediated coupling between the domain walls and demonstrated how the domain wall
dynamics is modified by the current passing through a pair of the domain walls in the wire.

Then we consider ferromagnetic nanorings with two domain walls.
In this case the DW dynamics is also affected by the coupling between the walls. Besides this,
there appear additional contributions to the gauge potential related to the ring geometry
of the ferromagnetic state and to the magnetic flux through the ring.
In the case of a relatively large ring with two domain walls, the problem can be solved
as for the wire with the periodic condition for eigenfunctions, resulting
in the quantization of values of the wavevector $k$ along the ring.
%However there are first a set of questions which need to be answered
%which have not yet been addressed in the literature. Real ferromagnetic ring systems do not in fact
%have a universal magnetization direction and can only locally be considered ferromagnetic.
%This complicates their behaviour in the ways we have discussed.

\begin{acknowledgement}
This work is supported by DFG under Grant No. SPP
1165 and partly by the FCT Grant PTDC/FIS/70843/2006 in
Portugal and by Polish Ministry of Science and Higher Education as a
research project in years 2007 -- 2010.
\end{acknowledgement}

\bibliographystyle{pss}
\bibliography{reportreferences}

\providecommand{\WileyBibTextsc}{}
\let\textsc\WileyBibTextsc
\providecommand{\othercit}{}
\providecommand{\jr}[1]{#1}
\providecommand{\etal}{~et~al.}


\begin{thebibliography}{[10]}

\bibitem{yamanouchi04}% article
 \textsc{M.~Yamanouchi},  \textsc{D.~Chiba},  \textsc{F.~Matsukura},  and
  \textsc{H.~Ohno},
 \jr{Nature} \textbf{428}, 539 (2004).


\bibitem{yamaguchi04}% article
 \textsc{A.~Yamaguchi},  \textsc{T.~Ono},  \textsc{S.~Nasu},
  \textsc{K.~Miyake},  \textsc{K.~Mibu},  and  \textsc{T.~Shinjo},
 \jr{Phys.~Rev.~Lett.} \textbf{92}, 077205 (2004).


\bibitem{parkin08}% article
 \textsc{S.~Parkin},  \textsc{M.~Hayashi},  and  \textsc{L.~Thomas},
 \jr{Science} \textbf{320}, 190 (2008).


\bibitem{klaui08}% article
 \textsc{M.~Klaeui},
 \jr{J.~Phys.~Condens.~Matter} \textbf{20}, 313001 (2008).


\bibitem{ebels00}% article
 \textsc{U.~Ebels},  \textsc{A.~Radulescu},  \textsc{Y.~Henry},
  \textsc{L.~Piraux}, ,  and  \textsc{K.~Ounadjela},
 \jr{Phys.~Rev.~Lett.} \textbf{84}, 983 (2000).


\bibitem{chopra02}% article
 \textsc{H.~Chopra} and  \textsc{S.~Hua},
 \jr{Phys.~Rev.~B} \textbf{66}, 020403 (2002).


\bibitem{ruster03}% article
 \textsc{C.~R{\"u}ster},  \textsc{T.~Borzenko},  \textsc{C.~Gould},
  \textsc{G.~Schmidt},  \textsc{L.~Molenkamp},  \textsc{X.~Liu},
  \textsc{T.~Wojtowicz},  \textsc{J.~Furdyna},  \textsc{Z.~Yu},  and
  \textsc{M.~Flatt\'e},
 \jr{Phys.~Rev.~Lett.} \textbf{91}, 216602 (2003).


\bibitem{vaz07}% article
 \textsc{C.~Vaz},  \textsc{T.~Hayward},  \textsc{J.~Llandro},
  \textsc{F.~Schackert},  \textsc{D.~Morecroft},  \textsc{J.\,A.\,C. Bland},
  \textsc{M.~Kl\"aui},  \textsc{M.~Laufenberg},  \textsc{D.~Backes},
  \textsc{U.~R\"udiger},  \textsc{F.\,J. Castano},  \textsc{C.\,A. Ross},
  \textsc{L.\,J. Heyderman},  \textsc{F.~Nolting},  \textsc{A.~Locatelli},
  \textsc{G.~Faini},  \textsc{S.~Cherifi},  and
  \textsc{W.~Wernsdorfer}\iffalse Ferromagnetic nanorings\fi,
 \jr{J.~Phys.~Condens.~Matter} \textbf{19}, 255207 (2007).


\bibitem{klaeui08}% article
 \textsc{M.~Kl\"aui}\iffalse Head-to-head domain walls in magnetic
  nanostructures\fi,
 \jr{J.~Phys.~Condens.~Matter} \textbf{20}, 313001 (2008).


\bibitem{sedlmayr09}% article
 \textsc{N.~Sedlmayr},  \textsc{V.\,K. Dugaev},  and
  \textsc{J.~Berakdar}\iffalse Current-induced interactions of multiple domain
  walls in magnetic quantum wires\fi,
 \jr{Phys. Rev. B} \textbf{79}, 174422 (2009).


\bibitem{sedlmayr10}% article
 \textsc{N.~Sedlmayr} and  \textsc{\emph{et al.}}\iffalse Spin and charge
  transport through non-collinear magnetic nanowires\fi,
 \jr{J. Magn. Magn. Mater} \textbf{322}, 1419 (2010).


\bibitem{sedlmayr08}% article
 \textsc{N.~Sedlmayr} and  \textsc{J.~Berakdar}\iffalse Transport properties of
  an interacting quantum dot with a non-uniform magnetization\fi,
 \jr{Eur. Phys. Lett.} \textbf{88}, 57003 (2008).


\bibitem{aharonov59}% article
 \textsc{Y.~Aharonov} and  \textsc{D.~Bohm}\iffalse Significance of
  electromagnetic potentials in the quantum theory\fi,
 \jr{Phys. Rev.} \textbf{115}(3), 485--491 (1959).


\bibitem{webb85}% article
 \textsc{R.\,A. Webb},  \textsc{S.~Washburn},  \textsc{C.\,P. Umbach},  and
  \textsc{R.\,B. Laibowitz}\iffalse Observation of $he$ aharonov-bohm
  oscillations in normal-metal rings\fi,
 \jr{Phys. Rev. Lett.} \textbf{54}(25), 2696--2699 (1985).


\bibitem{buettiker83}% article
 \textsc{M.~B\"uttiker},  \textsc{Y.~Imry},  and  \textsc{R.~Landauer}\iffalse
  Josephson behavior in small normal one-dimensional rings\fi,
 \jr{Physics Letters A} \textbf{96}(7), 365--367 (1983).


\bibitem{levy90}% article
 \textsc{L.\,P. L\'evy},  \textsc{G.~Dolan},  \textsc{J.~Dunsmuir},  and
  \textsc{H.~Bouchiat}\iffalse Magnetization of mesoscopic copper rings:
  Evidence for persistent currents\fi,
 \jr{Phys. Rev. Lett.} \textbf{64}(17), 2074--2077 (1990).


\bibitem{gefen84}% article
 \textsc{Y.~Gefen},  \textsc{Y.~Imry},  and  \textsc{M.\,Y. Azbel}\iffalse
  Quantum oscillations and the aharonov-bohm effect for parallel resistors\fi,
 \jr{Phys. Rev. Lett.} \textbf{52}(2), 129--132 (1984).


\bibitem{slonczewski96}% article
 \textsc{J.~Slonczewski},
 \jr{J. Magn. Magn. Mater.} \textbf{L1}, 159 (1996).


\bibitem{berger96}% article
 \textsc{L.~Berger},
 \jr{Phys.~Rev.~B} \textbf{54}, 9353 (1996).


\bibitem{barnas05}% article
 \textsc{J.~Barna\'s},  \textsc{A.~Fert},  \textsc{M.~Gmitra},
  \textsc{I.~Weymann},  and  \textsc{V.~Dugaev},
 \jr{Phys.~Rev.~B} \textbf{72}, 024426 (2005).


\bibitem{slonczewski89}% article
 \textsc{J.~Slonczewski},
 \jr{Phys.~Rev.~B} \textbf{39}, 6995 (1989).


\bibitem{theodonis06}% article
 \textsc{I.~Theodonis},  \textsc{N.~Kioussis},  \textsc{A.~Kalitsov},
  \textsc{M.~Chshiev},  and  \textsc{W.~Butler},
 \jr{Phys.~Rev.~Lett.} \textbf{97}, 237205 (2006).


\bibitem{kasai02}% article
 \textsc{S.~Kasai},  \textsc{T.~Niiyama},  \textsc{E.~Saitoh},  and
  \textsc{H.~Miyajima}\iffalse aharonov-bohm oscillation of resistance observed
  in a ferromagnetic fe-ni nanoring\fi,
 \jr{Applied Physics Letters} \textbf{81}, 316 (2002).


\bibitem{sekiguchi08}% article
 \textsc{K.~Sekiguchi},  \textsc{A.~Yamaguchi},  \textsc{H.~Miyajima},  and
  \textsc{A.~Hirohata}\iffalse Effect of ferromagnetism on ab oscillations in a
  normal-metal ring\fi,
 \jr{Phys. Rev. B} \textbf{77}, 140401 (2008).


\bibitem{klaeui03}% article
 \textsc{M.~Kl\"aui},  \textsc{C.\,A.\,F. Vaz},  \textsc{J.~Rothman},
  \textsc{J.\,A.\,C. Bland},  \textsc{W.~Wernsdorfer},  \textsc{G.~Faini},  and
   \textsc{E.~Cambril}\iffalse Domain wall pinning in narrow ferromagnetic ring
  structures probed by magnetoresistance measurements\fi,
 \jr{Phys. Rev. Lett.} \textbf{90}(9), 097202 (2003).


\bibitem{lai03}% article
 \textsc{M.\,F. Lai},  \textsc{Z.\,H. Wei},  \textsc{C.\,R. Chang},
  \textsc{J.\,C. Wu},  \textsc{J.\,H. Kuo},  and  \textsc{J.\,Y. Lai}\iffalse
  Influence of vortex domain walls on magnetoresistance signals in permalloy
  rings\fi,
 \jr{Phys. Rev. B} \textbf{67}(10), 104419 (2003).


\bibitem{jain08}% article
 \textsc{S.~Jain},  \textsc{C.\,C. Wang},  and  \textsc{A.\,O. Adeyeye},
 \jr{Nanotechnology} \textbf{19}, 085302 (2008).


\bibitem{tan96}% article
 \textsc{W.\,C. Tan} and  \textsc{J.\,C. Inkson},
 \jr{Semicond. Sci. Technol.} \textbf{11}, 1635 (1996).


\bibitem{kim99}% article
 \textsc{N.~Kim},  \textsc{G.~Ihm},  \textsc{H.\,S. Sim},  and  \textsc{K.\,J.
  Chang}\iffalse Electronic structure of a magnetic quantum ring\fi,
 \jr{Phys. Rev. B} \textbf{60}(12), 8767--8772 (1999).


\bibitem{tatara01}% article
 \textsc{G.~Tatara} and  \textsc{B.~Barbara}\iffalse Ferromagnetism\char39{}s
  affect on the aharonov-bohm effect\fi,
 \jr{Phys. Rev. B} \textbf{64}(17), 172408 (2001).


\bibitem{tatara04}% article
 \textsc{G.~Tatara},  \textsc{H.~Kohno},  \textsc{E.~Bonet},  and
  \textsc{B.~Barbara}\iffalse Aharonov-bohm oscillation in a ferromagnetic
  ring\fi,
 \jr{Phys. Rev. B} \textbf{69}(5), 054420 (2004).


\bibitem{loss90}% article
 \textsc{D.~Loss},  \textsc{P.~Goldbart},  and  \textsc{A.\,V.
  Balatsky}\iffalse Berry\char39{}s phase and persistent charge and spin
  currents in textured mesoscopic rings\fi,
 \jr{Phys. Rev. Lett.} \textbf{65}(13), 1655--1658 (1990).


\bibitem{meier07}% article
 \textsc{G.~Meier},  \textsc{M.~Bolte},  \textsc{.~Merkt},
  \textsc{B.~Kr\"uger},  and  \textsc{D.~Pfannkuche}\iffalse Current-induced
  domain-wall motion in permalloy semi rings\fi,
 \jr{J.~Mag.~Magnetic Materials} \textbf{316}, e966 (2007).


\bibitem{bruno05}% article
 \textsc{P.~Bruno} and  \textsc{V.\,K. Dugaev},
 \jr{Phys. Rev. B} \textbf{72}, 241302(R) (2005).


\bibitem{dugaev07}% article
 \textsc{V.\,K. Dugaev} and  \textsc{P.~Bruno},
 \jr{Phys. Rev. B} \textbf{75}, 201301(R) (2007).


\bibitem{kor}% article
 \textsc{V.~Korenman},  \textsc{J.~Murray},  and  \textsc{R.~Prange},
 \jr{PRB} \textbf{16}, 4032 (1977).


\bibitem{tatara}% article
 \textsc{G.~Tatara} and  \textsc{H.~Fukuyama},
 \jr{PRL} \textbf{78}, 3773 (1997).


\bibitem{dugaev1}% article
 \textsc{V.~Dugaev},  \textsc{J.~Barna\'s},  \textsc{A.~\L{}usakowski},  and
  \textsc{L.~Turski},
 \jr{Phys.~Rev.~B} \textbf{65}, 224419 (2002).


\bibitem{ijmp}% article
 \textsc{V.~Dugaev},  \textsc{V.~Vieira},  \textsc{P.~Sacramento},
  \textsc{J.~Barna\'s},  \textsc{M.~Ara\'ujo},  and  \textsc{J.~Berakdar},
 \jr{Int. J. Mod. Phys.} \textbf{21}, 1659 (2007).


\bibitem{dugaev3}% article
 \textsc{V.~Dugaev},  \textsc{J.~Berakdar},  and  \textsc{J.~Barna\'s},
 \jr{Phys.~Rev.~B} \textbf{86}, 104434 (2003).


\bibitem{sugawara08}% article
 \textsc{A.~Sugawara},  \textsc{H.~Kasai},  \textsc{A.~Tonomura},
  \textsc{P.~Brown},  \textsc{R.~Campion},  \textsc{K.~Edmonds},
  \textsc{B.~Gallagher},  \textsc{J.~Zeman},  and  \textsc{T.~Jungwirth},
 \jr{Phys.~Rev.~Lett.} \textbf{100}, 0477202 (2008).


\bibitem{claessen}% article
 \textsc{J.~Sch{\"a}fer},  \textsc{C.~Blumenstein},  \textsc{S.~Meyer},
  \textsc{M.~Wisniewski},  and  \textsc{R.~Claessen},
 \jr{Phys.~Rev.~Lett.} \textbf{101}, 236802 (2008).


\bibitem{bruno99}% article
 \textsc{P.~Bruno},
 \jr{Phys.~Rev.~Lett.} \textbf{83}, 2425 (1999).


\bibitem{pietzsch00}% article
 \textsc{O.~Pietzsch},  \textsc{A.~Kubetzka},  \textsc{M.~Bode},  and
  \textsc{R.~Wiesendanger},
 \jr{Phys.~Rev.~Lett.} \textbf{84}, 5212 (2000).


\bibitem{jb_du}% article
 \textsc{M.~Ara\'ujo},  \textsc{V.~Dugaev},  \textsc{V.~Vieira},
  \textsc{J.~Berakdar},  and  \textsc{J.~Barna\'s},
 \jr{Phys.~Rev.~B} \textbf{74}, 224429 (2006).


\bibitem{jb_du2}% article
 \textsc{V.~Dugaev},  \textsc{V.~Vieira},  \textsc{P.~Sacramento},
  \textsc{J.~Barna\'s},  \textsc{M.~Ara\'ujo},  and  \textsc{J.~Berakdar},
 \jr{Phys.~Rev.~B} \textbf{74}, 054403 (2006).


\bibitem{jb_du3}% article
 \textsc{V.~Dugaev},  \textsc{J.~Berakdar},  and  \textsc{J.~Barna\'s},
 \jr{Phys.~Rev.~Lett.} \textbf{96}, 047208 (2006).


\bibitem{jb_du4}% article
 \textsc{V.~Dugaev},  \textsc{J.~Barna\'s},  and  \textsc{J.~Berakdar},
 \jr{Phys.~Rev.~B} \textbf{71}, 024430 (2005).


\bibitem{dugaev2}% article
 \textsc{V.~Dugaev},  \textsc{J.~Barna\'s},  and  \textsc{J.~Berakdar},
 \jr{J. Phys. A} \textbf{36}, 9263 (2003).


\bibitem{meijer02}% article
 \textsc{F.\,E. Meijer},  \textsc{A.\,F. Morpurgo},  and  \textsc{T.\,M.
  Klapwijk}\iffalse One-dimensional ring in the presence of rashba spin-orbit
  interaction: Derivation of the correct hamiltonian\fi,
 \jr{Phys. Rev. B} \textbf{66}(3), 033107 (2002).


\end{thebibliography}

\end{document}